# Building a Chatbot on a Closed Domain using RASA

Khang Nhut Lam
Can Tho University
Vietnam
lnkhang@ctu.edu.vn

Nam Nhat Le
Can Tho University
Vietnam
lenhatnam10b5@gmail.com

Jugal Kalita
University of Colorado
USA
jkalita@uccs.edu.vn

## ABSTRACT
In this study, we build a chatbot system in a closed domain with the RASA framework, using several models such as SVM for classifying intents, CRF for extracting entities and LSTM for predicting action. To improve responses from the bot, the kNN algorithm is used to transform false entities extracted into true entities. The knowledge domain of our chatbot is about the College of Information and Communication Technology of Can Tho University, Vietnam. We manually construct a chatbot corpus with 19 intents, 441 sentence patterns of intents, 253 entities and 133 stories. Experiment results show that the bot responds well to relevant questions.

## CCS Concepts
• **Information systems ~ Information retrieval ~ Retrieval tasks and goals ~ Question answering**.

## Keywords
chatbot; Rasa; SVM; CRF; LSTM; kNN.

## 1. INTRODUCTION
In recent years, the concepts of virtual assistants or chatbots has become commonplace. The leading technology corporations have officially released their virtual assistants such as Cortana of Microsoft, Siri of Apple, and Google Assistant of Google. There are several available tools for building chatbots such as Chatfuel, Messnow, ChattyPeople and ManyChat. However, these tools may have advertisements and do not support many languages, including the Vietnamese language.

A chatbot can be constructed using the following approaches: using retrieval-based or generative models, supporting short or long conversations, and in closed or open domains. Bots created using the generative model can answer questions which are not in the training dataset, but these answers might be in wrong syntax or have misspelling; whereas bots created using the retrieval-based model can respond answers with correct grammar and spelling, only for questions in the training dataset. The lengths of conversations also affect bots. The longer the text conversations are, the harder it is to construct the bots. Bots constructed in an open domain require a large amount of knowledge in order to answer unrestricted questions; whereas bots built in a closed domain can answer questions in that specific domain. Although there are a variety of methods for constructing a chatbot, each method for building chatbot systems needs to handle the following issues: classifying, determining, and extracting intents that users express. In addition, a smart chatbot needs to handle acronyms and misspelled words.

In this paper, we present our work on building a chatbot system for students at the College of Information and Communication Technology (CICT) of Can Tho University in Vietnam. The chatbot system is required to answer introductory questions about CICT, including programs and staff, academic regulations, study plans and classes. Therefore, our chatbot is constructed using a retrieval-based method in a closed domain with short conversations. The remainder of this paper is organized as follows. In Section 2, we present related work. Section 3 describes approaches we propose to construct the chatbot system. Experiments and results are discussed in Section 4. Section 5 concludes the paper.

## 2. RELATED WORK
Tsung-Hsien et al. [1] construct a dialogue system using an end-to-end task-oriented dialogue system and a seq2seq model to train on a dataset gathered from the Wizard-of-Oz novel. The results show that the model has a relatively high accuracy, and it is able to converse with humans quite naturally with an average BLEU score of 0.23. Guo et al. [2] develop a chatbot on Tensorflow and MXNet frameworks using the seq2seq model. A dataset with emojis consisting of the Cornell movie Dialog Corpus with 221,282 question-answer pairs and the Twitter chat corpus with 377,265 question-answer pairs are used to train the model. Their system can capture simple entities, but most of the responses are quite general. To help make bot responses more appropriate, Dhyani and Kumar [3] construct an assistant conversational agent using bidirectional recurrent neural networks and an attention model. They train the chatbot model on a Reddit dataset. The perplexity and the BLEU score of their model are 56.10 and 30.16, respectively.

Segura et al. [4] develop a social chatbot, called Chatbol, on the football domain related to the "La Liga" football league. A main component of Chatbol is an NLU block trained to extract intents and entities from questions. The entities extracted are used to query Wikidata to obtain information and respond to users. The training dataset is created by extracting dialogues about football on television channels and football segments from the OpenSubtitles dataset. Chatbol is built on the RASA framework. Chatbol produces relatively good results with 72% of responses being are relevant to user questions. Muangkammuen et al. [5] introduce a chatbot system, named Thai-FAQ, that can automatically answer questions for customers. This chatbot is constructed using an LSTM model. Experiments show that the system recognizes 86.36% questions and responds with appropriate answers with 93.2% accuracy.

Ming et al. [6] present a method to build a chatbot for the elderly. Data extracted from the MHMC chitchat dataset is used for training. An LSTM-based multilayer embedding model is used to extract semantic information, and Euclidean distance is used to calculate and select a relevant question and answer from the dataset. The model achieves 79.96% accuracy for the first answer. In a similar work, Tascini [7] builds a chatbot to assist the elderly using a Deep Belief Network [8]. The author trains the model on a variety of corpora such as the Ubuntu corpus, Microsoft Research Social Media Conversation corpus and Cornell Movie Dialog corpus. Experiments show that the system can learn by itself through conversations. Kataria et al. [9] have developed a depression reduction chatbot system, called Bot-Autonomous Emotional Support. The bot is built using an encoder-decoder model with 3 layers of LSTMs on the Tensorflow framework. The system is able to learn dialogues by itself in order to provide better responses.

Personalizing a dialogue system requires sufficient information from users. Nguyen et al. [10] use a seq2seq model to build an open domain dialogue system that mimics characters from popular TV shows such as Barney from "How I Met Your Mother" and Sheldon from "The Big Bang Theory". Their system achieves quite satisfactory with more than 50% human judges believing that they are not chatting with bots. In another related work, Li et al. [11] propose a persona-based model, which is a combination of the seq2seq model, Speaker Model and Speaker-Addressee model, to construct a chatbot. A dataset consisting of 25 million Twitter conversations is used for training. The model gives better results than the traditional seq2seq model in terms of BLEU scores and judgment on the speaker's personality to give appropriate responses. In particular, the BLEU scores yield an increase of 21% in the maximum likelihood setting and 11.7% for the maximum mutual information setting.

Iulian et al. [12] use a deep reinforcement learning approach to develop a chat bot, called MILABOT. This bot is able to interact with humans via both text and speech. MILABOT is a combination of a natural language generation model and a retrieval model, including reinforcement learning, template-based and bag-of-word approaches, and seq2seq and latent variable neural network models. The authors claim that the system has better performance than other existing systems.

In Vietnamese, Vu [13] has built a dialogue system using seq2seq and LSTM models. The system is trained on the OpenSubtitles 2016 dataset. The author has not evaluated the system, but he claims that the model results are good.

Our chatbot system is built on the RASA framework. Support Vector Machines (SVM) and Conditional Random Fields (CRF) are used to classify intents and extract entities, respectively. The k-Nearest Neighbor (kNN) algorithm is used to predict correct entities and an LSTM model is used to manage the dialogue.

## 3. PROPOSED APPROACH

Kompella [1] describes a chatbot architecture with three main components: Natural Language Understanding (NLU), Dialog Management (DM) and Message Generator (MG). The NLU component determines an intent and extracts entities from a user request. The extracted intent and entities are fed into the DM component, which predicts the next action based on the trained stories. The DM component is also responsible for requesting data from other systems through API. The MG component extracts a relevant response regarding the action identified in the DM from the pre-defined templates. We use this chatbot system architecture to construct our chatbot system.

### 3.1 Generate the NLU Data

We construct a chatbot system in a closed domain about CICT. The NLU data consists of intents, including names and sentence patterns. Each intent may or may not have entities, each of which includes values and names. Figure 1 presents an intent named "XinChao" (means "Greeting") and its sentence patterns without entities.

```
## intent:XinChao
- xin chào
- chào
- alo
- hello
- hi
- chào bot
```

**Figure 1. An example of an intent without entities**

Figure 2 shows an example of an intent named "dinhNghia" (means "definition"). This intent has sentence patterns with entities. Each entity has a value and a name. In particular, the first sentence pattern of this intent has an entity value of "chương trình đào tạo" (means "program") and an entity name of "dn". We construct a total of 19 intents with 441 intent sentence patterns, 6 entity names and 253 entity values.

```
## intent:dinhNghia
- tôi muốn biết [chương trình đào tạo](dn) là gì
- [kế hoạch học tập](dn) là như thế nào
- [học phần](dn) là cái gì
- [học phần tiên quyết](dn) là sao
- cho tôi biết [học phần bắt buộc](dn) là gì
- [lớp chuyên ngành](dn) là gì
```

**Figure 2. Example of an intent with entities**

### 3.2 Create Relevant Responses

For each question intent, we design a variety of pattern answers to make the bot agile, not too stereotyped and boring. With the example response patterns shown in Figure 3, the bot can perform an action "utter_xinChao" using one of three answer patterns provided.

```
templates:
  utter_xinChao:
    - text: Hey! Chào bạn <3 !
    - text: Chào bạn !
    - text: Xin chào !
```

**Figure 3. Example of answer patterns**

### 3.3 Build Dialogue Data

Dialogue data, stories or sample conversations represent a conversation and associated information between a user and a chatbot system from start to finish. Based on these conversations, the chatbot system predicts the context and takes the next action.

---

[1] https://towardsdatascience.com/architecture-overview-of-a-conversational-ai-chat-bot-4ef3dfefd52e

For each intent, the bot responds with a relevant answer corresponding to the context of the dialogue. A large size of dialogue data helps the bot predict better and perform actions more accurately. Figure 4 shows an example of dialogue data. In Figure 4, the *utter_continue* is the previous action, *dinhNghia* is an intent, *dn* is an entity, *action_dn* is the next action.

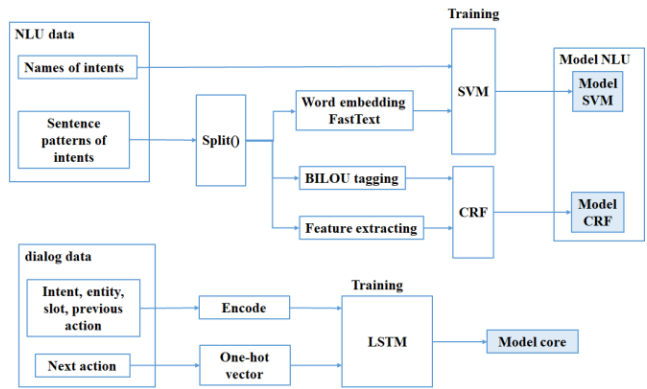

**Figure 4. Example of dialogue data**

### 3.4 Train the Model

RASA framework[2] is used to construct the chatbot system. The process to train the model is presented in Figure 5. We use the SVM implementation in Scikit-learn with parameters configured by RASA for classifying intents. The CRF and LSTM implementations, also provided by RASA, are used to extract entities and predict the next action, respectively.

**Figure 5. The model for constructing a chatbot**

As mentioned earlier, the NLU data consists of intents with names and intent sentence patterns. Each name of intent, which is considered a class, is converted to numbers. Sentence patterns of intents are tokenized by spaces and converted to vectors using fastText[3]. Then, these data are trained to classify intents by the SVM model with the GridSearchCV tool using the *C* parameter in the set of 6 values [1, 2, 5, 10, 20, 100]. In addition, after tokenizing sentence patterns, they are labeled by the BILOU technique and extracted features; then they are fed to the CRF model for training to predict entities. The CRF model with $max\_iterations$ (the number the maximum iterations for the optimization algorithm) are 50, and L1_c and L2_c (which are the parameters for the loss functions) are equal to 0.1.

---

[2] https://rasa.com/

[3] https://fasttext.cc/

The BILOU technique uses a B tag for the beginning of the entity, an I tag for the middle entity, an L for the end of entity, an O for the non-entity and a U for entity having a single word. An example of application of the BILOU technique is shown in Fig 6.

**Figure 6. An example of using BILOU tags**

In the training step, RASA uses the *max_history* (parameter specifying the number of previous states to include in the futurization) of 5. Input data including intents, entities, slots and previous actions are converted to a binary vector of the size of the sum of intents, entities, slots, and previous actions. The next action is converted to a vector using the one-hot vector method with the size of the total actions. Finally, data are fed to the LSTM model to train to learn the next action.

### 3.5 Improve the Model

The process of extracting entities of the CRF model produces good results only for input containing the correct entities. To help the system extract false entities, we generate new words such that these words are misspellings or have no diacritics written above or below the vowels, as shown in Figure 7.

**Figure 7. An example of generating incorrect words**

We improve the model by applying the kNN algorithm to convert incorrect entities into correct entities. The purpose of this process is to help the bot understand and respond with relevant answers even in case users mistyped. In other words, we try to assign a correct label of a correct entity to an incorrect entity. First, we build a training dataset by generating incorrect entities. Second, we create feature vectors for these incorrect entities. Next, these feature vectors with their labels are trained using the kNN algorithm and the Euclidean distance. In addition, regular expressions are used to process data to convert incorrect entities to correct entities.

## 4. EXPERIMENTS AND RESULTS

We run experiments on PCs with Ubuntu 18.04.3 LTS operating system, Intel core i3-4010U@1.7GHz CPU and 8G Ram. The data set includes 441 questions belonging to 19 intents (labels) such as "xinChao" (means "Greeting"), "gioiThieuKhoa" (means "College Introduction") and "HPTQ" (stands for "Học Phần Tiên Quyết" in Vietnamese and means "Prerequisite Subject" in English); 253 entities, 133 stories and 1,336 response actions. Since the entity prediction process uses the kNN algorithm, our bot can predict and provide relevant answers for misspelled questions. In addition, we use Google SpeechRecognition[4] to input voice.

The division of data for model evaluation is performed using StratifiedKFold method. StratifiedKFold divides the data into 10 sections including 9 sections used for training and the last one for testing. Figure 8 presents the results of predicting intents using SVM with a Confusion Matrix. Figure 8 shows that the system predicts 29 questions incorrectly. The questions on the diagonal of the intent confusion matrix have correct responses. Table 1 presents accuracies of the model using different kernels. The model achieves the best accuracy of 94.33% with the nonlinear kernel "rbf".

---

[4] https://cloud.google.com/speech-to-text

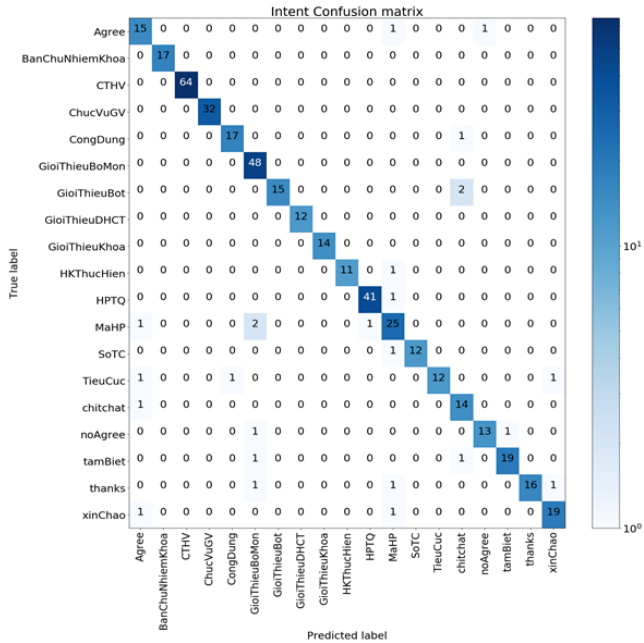

**Figure 8. The intent confusion matrix**

**Table 1. Accuracies of kernels**

| Kernel | linear | poly | sigmoid | rbf |
|---|---|---|---|---|
| Accuracy | 93,65 | 85,00 | 92,30 | 94,33 |

The predicting probability distribution chart, Figure 9, also shows that the questions with incorrect classification have a confidence value between 0.15 and 0.45.

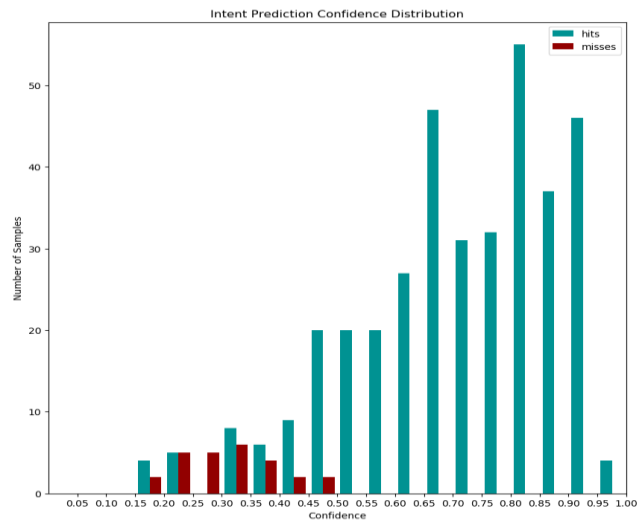

**Figure 9. The predicting probability distributions**

The prediction of entity labels by the CRF model achieves 95% accuracy. We also evaluate the results of bot responses. We build a test set consisting of 8 correct stories with 99 responses. The results show that there are 6 correct stories (75% accuracy) and 90 relevant responses (92.78%). The action confusion matrix is shown in Figure 10.

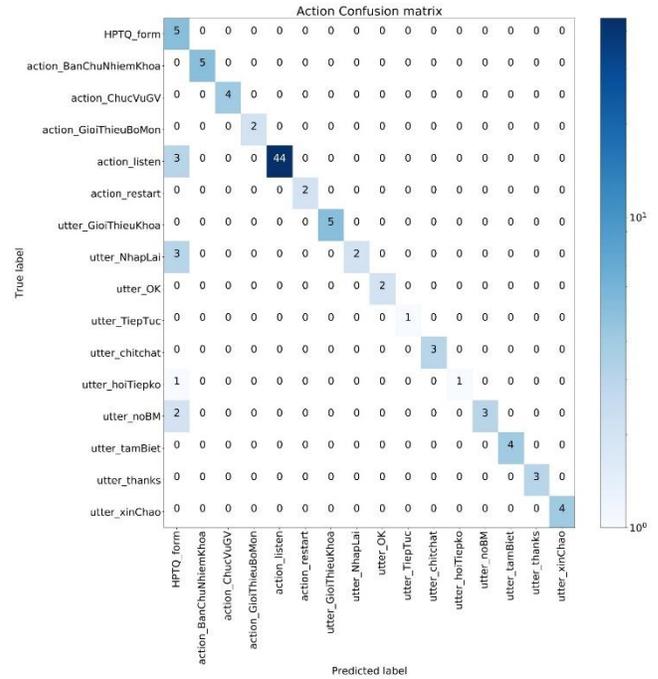

**Figure 10. Action confusion matrix**

We increase bot responsiveness by applying the kNN algorithm to recognize the wrong entities and convert them to correct entities. We perform experiments with different values of *k* to find a good value as shown in Table 2. With *k* =17, the model produces the best result.

**Table 2. The *k* value and the model's accuracy**

| k | 11 | 13 | 15 | 17 | 19 |
|---|---|---|---|---|---|
| Accuracy | 97,14 | 97,18 | 97,21 | 97,25 | 97,24 |

## 5. CONCLUSION

In this paper, we have built a closed domain chatbot system for the CICT on the RASA Framework. The system responds quite well with questions belonging to trained intents. Besides, the system can also understand and answer questions with moderate misspellings. A chatbot system that responds with answers extracted from the pre-built sentence dataset can overcome wrong syntax and incorrect spelling answers. However, the responses are sometimes unnatural and the system cannot answer questions outside of the training dataset. For future work, we will enlarge the training data and use a seq2seq model to help the system answer questions which are not in the training dataset.